\documentstyle[epsf]{mn}

\newif\ifAMStwofonts

\ifoldfss
  \ifCUPmtlplainloaded \else
    \NewTextAlphabet{textbfit} {cmbxti10} {}
    \NewTextAlphabet{textbfss} {cmssbx10} {}
    \NewMathAlphabet{mathbfit} {cmbxti10} {} 
    \NewMathAlphabet{mathbfss} {cmssbx10} {} 
  \fi
  \ifAMStwofonts
    \ifCUPmtlplainloaded \else
      \NewSymbolFont{upmath} {eurm10}
      \NewSymbolFont{AMSa} {msam10}
      \NewMathSymbol{\upi}     {0}{upmath}{19}
      \NewMathSymbol{\umu}     {0}{upmath}{16}
      \NewMathSymbol{\upartial}{0}{upmath}{40}
      \NewMathSymbol{\leqslant}{3}{AMSa}{36}
      \NewMathSymbol{\geqslant}{3}{AMSa}{3E}

      \let\geq=\geqslant 
    \fi
  \fi
\fi 

\ifnfssone
  \newmathalphabet{\mathit}
  \addtoversion{normal}{\mathit}{cmr}{m}{it}
  \addtoversion{bold}{\mathit}{cmr}{bx}{it}
  \newmathalphabet{\mathbfit} 
  \addtoversion{normal}{\mathbfit}{cmr}{bx}{it}
  \addtoversion{bold}{\mathbfit}{cmr}{bx}{it}
  \newmathalphabet{\mathbfss} 
  \addtoversion{normal}{\mathbfss}{cmss}{bx}{n}
  \addtoversion{bold}{\mathbfss}{cmss}{bx}{n}
  \ifAMStwofonts
    \ifCUPmtlplainloaded \else
      \UseAMStwoboldmath
      \makeatletter
      \new@mathgroup\upmath@group
      \define@mathgroup\mv@normal\upmath@group{eur}{m}{n}
      \define@mathgroup\mv@bold\upmath@group{eur}{b}{n}
      \edef\UPM{\hexnumber\upmath@group}
      \new@mathgroup\amsa@group
      \define@mathgroup\mv@normal\amsa@group{msa}{m}{n}
      \define@mathgroup\mv@bold\amsa@group{msa}{m}{n}
      \edef\AMSa{\hexnumber\amsa@group}
      \makeatother
      \mathchardef\upi="0\UPM19
      \mathchardef\umu="0\UPM16
      \mathchardef\upartial="0\UPM40
      \mathchardef\leqslant="3\AMSa36
      \mathchardef\geqslant="3\AMSa3E

      \let\geq=\geqslant 
    \fi
  \fi
\fi 

\ifnfsstwo
  \DeclareMathAlphabet{\mathbfit}{OT1}{cmr}{bx}{it}
  \SetMathAlphabet\mathbfit{bold}{OT1}{cmr}{bx}{it}
  \DeclareMathAlphabet{\mathbfss}{OT1}{cmss}{bx}{n}
  \SetMathAlphabet\mathbfss{bold}{OT1}{cmss}{bx}{n}
  \ifAMStwofonts
    \ifCUPmtlplainloaded \else
      \DeclareSymbolFont{UPM}{U}{eur}{m}{n}
      \SetSymbolFont{UPM}{bold}{U}{eur}{b}{n}
      \DeclareSymbolFont{AMSa}{U}{msa}{m}{n}
      \DeclareMathSymbol{\upi}{0}{UPM}{"19}
      \DeclareMathSymbol{\umu}{0}{UPM}{"16}
      \DeclareMathSymbol{\upartial}{0}{UPM}{"40}
      \DeclareMathSymbol{\leqslant}{3}{AMSa}{"36}
      \DeclareMathSymbol{\geqslant}{3}{AMSa}{"3E}

      \let\geq=\geqslant 
    \fi
  \fi
\fi 

\ifCUPmtlplainloaded \else
  \ifAMStwofonts \else 
    \def\upi{\pi}
    \def\umu{\mu}
    \def\upartial{\partial}
  \fi
\fi

\title[A new PMS association]
  {The discovery of a low mass, pre-main-sequence stellar association around $\gamma$-Velorum}
\author[M. Pozzo et al.]
  {M. Pozzo, 
  R.D.~Jeffries, T.~Naylor, E. J.~Totten\thanks{Visiting astronomer at the
  Cerro Tololo Interamerican Observatory}, 
  S.~Harmer\thanks{Nuffield Foundation 
Undergraduate Research Bursar (NUF-URB98)}, 
M.~Kenyon\thanks{Present address, University College, Gower St., London
  WC1E 6BT}\\
  Department of Physics, Keele University, Keele, Staffordshire, ST5 5BG, UK}
\date{Accepted 0000 Received 0000}
\pagerange{\pageref{firstpage}--\pageref{lastpage}}
\pubyear{0000}

\newcommand{\kms}{\,km\,s$^{-1}$}

\newcommand{\msun}{\,\mbox{$\mbox{M}_{\odot}$}}

\newcommand{\be}{\begin{equation}}
\newcommand{\ee}{\end{equation}}
\newcommand{\bd}{\begin{displaymath}}
\newcommand{\ed}{\end{displaymath}}

\begin{document}

\label{firstpage}
\maketitle

\begin{abstract}
We report the serendipitous discovery of a population of low mass,
pre-main sequence stars (PMS) in the direction of the Wolf-Rayet/O-star
binary system $\gamma^{2}$ Vel and the Vela OB2 association.  We argue
that $\gamma^{2}$ Vel and the low mass stars are truly associated, are
approximately coeval and that both are at distances between
360-490\,pc, disagreeing at the 2$\sigma$ level with the recent Hipparcos
parallax of $\gamma^{2}$ Vel, but consistent with older distance estimates.  
Our results clearly have implications
for the physical parameters of the $\gamma^{2}$ Vel system, but also
offer an exciting opportunity to investigate the influence of high mass
stars on the mass function and circumstellar disc lifetimes of their
lower mass PMS siblings.


\end{abstract}


\section{Introduction}

As little as 15 years ago it was believed that high mass stars formed
mainly in OB associations and that low mass stars formed mainly in T
associations. In the last few years it has become increasingly
recognized that the majority of low mass stars in the solar
neighbourhood are in fact likely to have formed in OB
associations. The mass function in OB associations may
match popular log-normal parameterisations of the general field population 
such as those proposed by Miller \& Scalo (1979), with many low
mass stars formed for every O/B star. The main reason for
this shift in paradigm is the discovery of numerous low mass pre-main
sequence (PMS) stars in young OB associations by virtue of their high levels
of X-ray activity (see for example Walter et al. 1994, 1999; Naylor \&
Fabian 1999, Preibisch \& Zinnecker 1999 and references
therein). Recently, some evidence has been found that these X-ray
selected groups can be quite small, concentrated around just one or two
high mass stars, which are themselves part of larger
associations. Examples include
$\beta$ Cru (Feigelson \& Lawson 1997), $\sigma$ Ori (Walter,
Wolk \& Sherry 1998) and $\eta$ Cha (Mamajek, Lawson \& Feigelson
1999).

The investigation of these low mass PMS stars is of prime importance in
establishing the form of the {\em initial} mass function. Although OB
associations are by definition unbound, they are young enough that
dynamical effects such as mass segregation and preferential evaporation
will not have occurred; and they are usually free of the heavy,
variable extinction that plagues observations of active star forming
regions. If the majority of low mass stars are born in OB associations
it is crucial to establish the influence that high mass
neighbours have on the formation and evolution of their lower mass
siblings. The winds and ionizing radiation of hot
stars could influence the mass function and circumstellar disc
lifetimes of the lower mass stars, with implications for angular
momentum evolution and planet formation.  These ideas have gained
currency with the discovery of evaporating discs around PMS stars
in the Orion nebula (McCaughrean \& O'Dell 1996) and theoretical
studies showing that discs could be ionized and evaporated by
the UV radiation fields of O stars (Johnstone, Hollenbach \& Bally
1998). In more extreme circumstances, nearby supernova explosions have
been invoked both as a means of terminating low mass star formation (Walter
et al. 1994) or triggering it (Preibisch \& Zinnecker 1999).

In this paper we report the discovery, by X-ray selection, of a low mass
stellar population that seems likely to be associated with the nearest
example of a Wolf-Rayet (WR) star, $\gamma^{2}$ Velorum
(HD 68273, HIP 39953, WR11). Like about
half of the $\simeq 200$ galactic WR stars known, it is a binary
system (WC8+O8) with an orbital period of 78.5 days and a massive,
interacting stellar wind. There is currently some controversy
concerning the distance to $\gamma^{2}$ Vel, which impacts upon the
deduced luminosities, masses and mass loss rates from the system. It is
important to get these parameters right because, as the nearest WR,
$\gamma^{2}$ Vel is an extreme test of stellar evolution models and
calibrates the absolute magnitudes of WR stars.  The Hipparcos
parallax yields a distance of $258^{+41}_{-31}$\,pc to $\gamma^{2}$ Vel
(Schaerer, Schmutz \& Grenon 1997, van der Hucht et al. 1997), in
marked contrast to previous distance estimates which place it at
350-450\,pc. The larger distance is in better agreement with the mean
distance to the Vela OB2 association ($410\pm12$\,pc), of which
$\gamma^{2}$ Vel is the most massive proper-motion member (de Zeeuw et
al. 1999). 

The presence of a low mass stellar association in the extreme
environment around $\gamma^{2}$ Velorum offers an excellent empirical
test of the possible influence of winds and ionizing radiation on low mass 
stars and their discs. Additionally it gives us a chance to measure the
distance to $\gamma^{2}$ Vel by matching PMS isochrones to the low mass
stars.

\section{X-Ray observations}

\begin{figure}
\vspace*{8.0cm}
\includegraphics{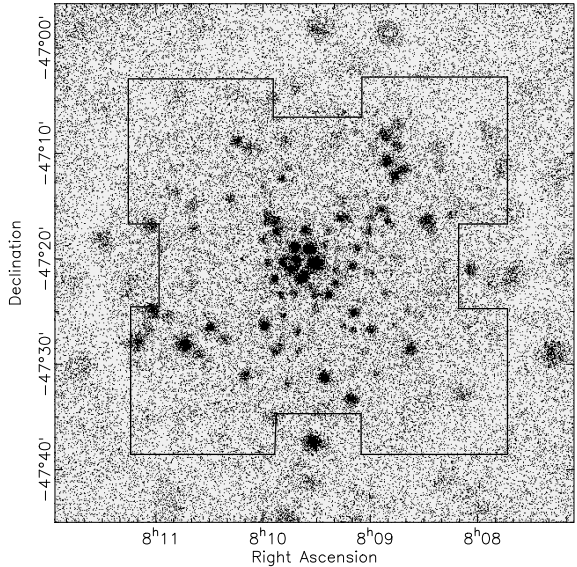}
\caption{{\em ROSAT} PSPC X-ray image of the region around $\gamma^{2}$
Vel. The greyscale is such that black represents $\geq 1$ photon per
5x5 arcsec pixel and the brightest source in the centre of the image is
$\gamma^{2}$ Vel. Another 108 significant X-ray sources have been found in this
image. The solid outline shows the location of our optical CCD survey.}
\end{figure}

We suspected the presence of a low mass association around $\gamma^{2}$
Vel from the large surrounding population of X-ray point sources seen
in {\em ROSAT} images, taken for a programme
to investigate its interacting stellar winds (see Willis, Schild \&
Stevens 1995 and Fig.1).
The X-ray observations of the WR were retrieved from the
{\em ROSAT} public archive and consisted of 10 Position Sensitive
Proportional Counter (PSPC) datasets and 2 datasets taken with the High
Resolution Imager. In this initial paper we discuss only the more
sensitive PSPC results, which contain the vast majority of the X-ray point sources.

We used the Starlink-distributed 
{\sc asterix} data reduction package in our X-ray analysis (Allan \& Vallance 1995).
The 10 PSPC datasets were sorted into 1$^{\circ}\times1^{\circ}$ images, selecting
pulse height channels 11 to 240 (approximately 0.1-2.4\,keV photons) 
and excluding times with anomalously high background rates. 
The images were centered on
$\gamma^{2}$ Vel (the brightest source in each dataset) and then
summed. The resulting image has dimensions of $720 \times 720$
5 arcsecond pixels and an effective on-axis exposure time of
25.3 ks. We used the Point Source Searching (PSS - Allan 1992)
algorithm to search for sources by
Cash-statistic maximisation. By assuming the background to be zero we
obtained a preliminary list 
of 104 X-ray sources which were masked out of the image. The masked
image was then patched and smoothed with a 75 arcsec FWHM gaussian to
create a background map. We then executed the PSS algorithm again and
found 109 sources above a pseudo-gaussian significance level of
4.5$\sigma$, which corresponds roughly to 1 spurious detection in the
X-ray image. The positions of the X-ray sources were corrected for errors in the
satellite aspect solution by comparing the optical and X-ray positions
of $\gamma^{2}$ Vel and four other bright stars from the CDS SIMBAD
database. We applied shifts of
4.5 arcsec in RA and 13.8 arcsec in Dec to the X-ray positions.

\section{Optical photometry}

\begin{figure}
\vspace*{15.0cm}
\includegraphics{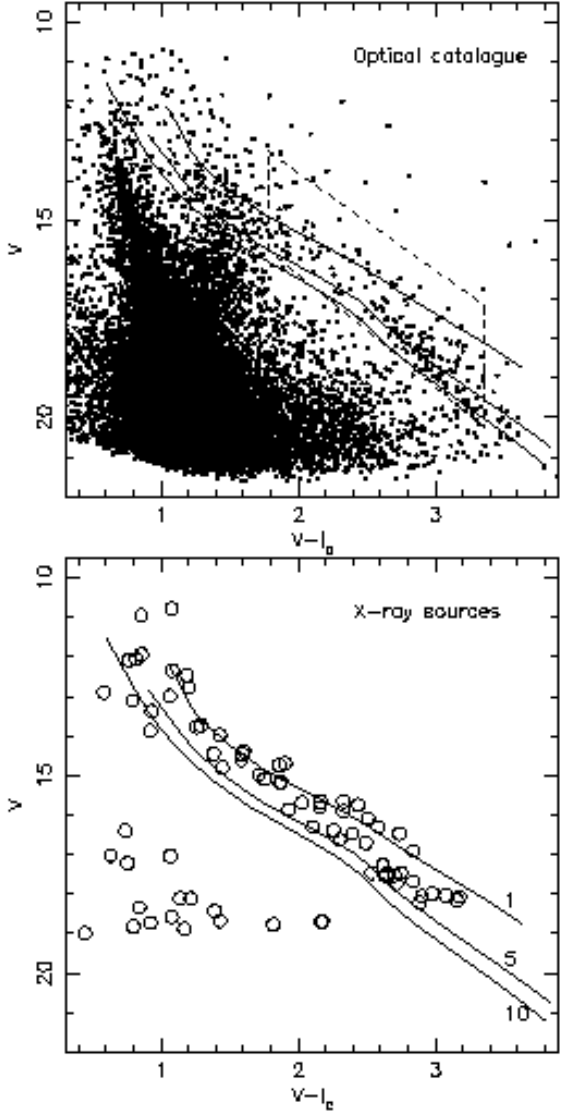}
\caption{(Top) CMD for the region around $\gamma^{2}$ Vel. The solid
lines in this and the plot below are isochrones at 1, 5 and 10\,Myr
calculated from the models of D'Antona \& Mazzitelli (1997). The dashed
box indicates the region from which PMS stars were selected for
Fig.3. (Bottom) The location of the X-ray source correlations in the
CMD. Eleven spurious correlations are expected and are expected to lie
in the clump at $V\simeq 19$, $V-I\simeq1$.}
\end{figure}

CCD photometry of the X-ray field was obtained on 8 February 1999 with
the 0.9-m telescope at the Cerro Tololo
Inter-American Observatory. A Tek 2048x2048 CCD was
used to give a 13.5x13.5 arcmin$^{2}$ field of view. 
Eight overlapping fields around $\gamma^{2}$-Vel were
surveyed in {\it BVI} with short (20,10,10s) and long (200,100,100s) 
exposures, together with five fields from Landolt (1992) to determine 
zeropoints, colour terms and extinction coefficients. Several standards
with $V-I_{c}>2.5$ were observed. Figure~1
illustrates the location of the fields around $\gamma^{2}$\,Vel. The external
accuracy of our photometry was determined to be 
around 0.02 mag on the $BVI_{\rm c}$ system.

Photometry was performed for each of the eight fields using an optimal
aperture algorithm (Naylor 1998; Totten et al. in preparation), and the results
combined to give an optical catalogue of sources. 
Astrometry was performed by comparison with star positions
in version A2.0 of the USNO catalogue (Monet 1998) and we estimate
positional accuracies of around 0.3 arcsec. The final catalogue
contains 20617 individual objects with their magnitudes, colours and
positions.  Figure 2a shows the $V$ versus $V-I_{\rm c}$ 
colour-magnitude diagram (CMD). The location of a possible
PMS low-mass star association is clearly visible above the bulk of the
background contamination. To this optical catalogue, we added the
positions of bright ($V<11$)
stars found in this region from the SIMBAD database.

To establish an appropriate cross-correlation  radius to use between
the X-ray and optical source lists we modelled the cumulative number of
X-ray sources that were correlated with an optical source with $V<19$
(see Jeffries, Thurston \& Pye 1997 for details). Assuming a uniform spread of
optical sources, we determined that there were 77 correlations (from 83
X-ray sources inside the CCD survey) within
10 arcsec of X-ray positions, that 66
of these would be true counterparts to X-ray
sources, 11 would be spurious correlations and 
that the $1\sigma$ X-ray error circle was 3.7 arcsecs.
Figure~2b shows 75 sources that have an optical counterpart within 10 arcsecs (another
two are bright stars without $V-I$ colours).
Clearly the X-ray emitting population coincides with the proposed PMS
population in the CMD. Indeed, if we were to consider just a subset of the
optical catalogue consisting of a broad strip containing all these PMS
sources, we would only expect 1 of these correlations to be spurious.

We have calculated the X-ray properties of this population and these
along with the HRI observations will be reported in a subsequent
publication. Briefly, the X-ray to bolometric flux ratio of these
objects lies in the range $10^{-5}$ to $10^{-2}$, broadly what we would
expect from a population of young PMS stars. The cut-off in the
PMS X-ray correlations at $V\simeq18$ is almost certainly due to the X-ray
sensitivity.  To be detected in X-rays, fainter objects would have to
have higher than feasible X-ray to bolometric flux ratios. However, it is
clear that the PMS sequence we have found extends down to the limits of
our optical survey at $V\sim20.5$.

\section{Discussion}
\begin{figure}
\vspace*{15.0cm}
\includegraphics{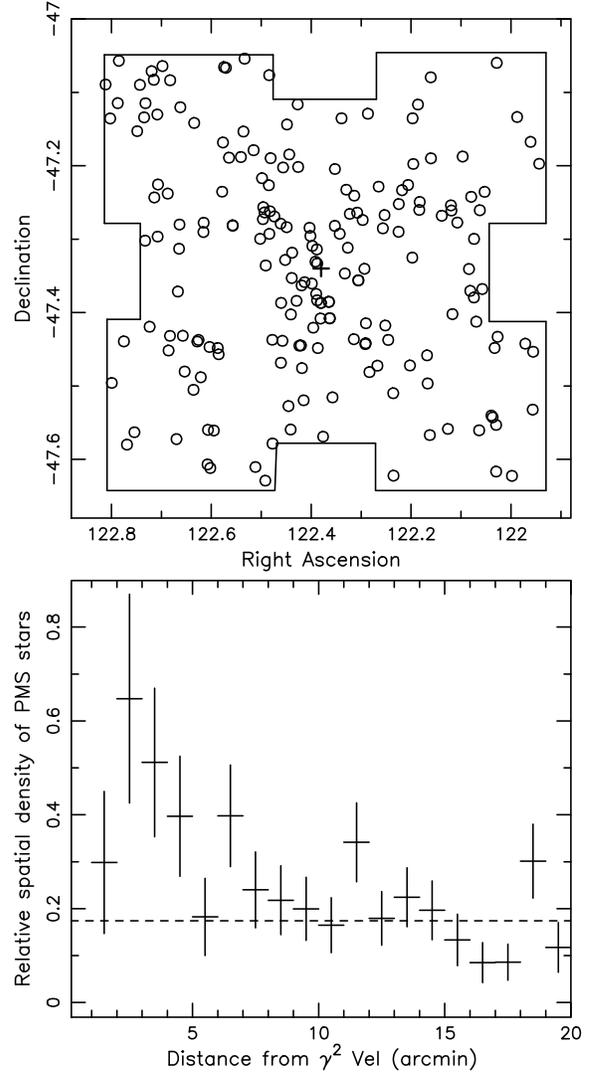}
\caption{(Top) The spatial distribution of PMS stars (from inside the
dashed box in Fig.2) around $\gamma^{2}$ Vel. A clumping towards
$\gamma^{2}$ Vel (marked with a cross) is apparent.
(Bottom) The relative spatial density of these PMS sources as a
function of distance from $\gamma^{2}$ Vel. This plot is normalized
using a background population to account for non-uniform coverage in
the CCD survey.}
\end{figure}

The appearance of Fig.2 should leave the reader in no doubt that we
have found a young and exceptionally rich population of low mass active
stars in the {\em direction} of $\gamma^{2}$ Vel. The central question to
be answered is whether these sources are physically close to
$\gamma^{2}$ Vel and/or whether they are background members of the Vela OB2
association. We can tackle this problem in a number of ways.

\subsection{Are the PMS stars and $\gamma^{2}$ Vel aligned?}

Figure 3a shows the spatial distribution of
PMS stars selected from the CMD 
in a strip enclosing the bulk of the X-ray sources (marked with
a dashed box in Fig.2). This box was chosen to avoid background
contamination.  We determined the radial distribution of these stars,
centred on $\gamma^{2}$ Vel. The
distribution is normalized using the radial distribution of background
stars with a similar $V$
magnitude range, but a colour range of 0.8$<V-I_{\rm c}<$1.7, 
under the assumption that the background
stars are uniformly distributed. 
The resulting radial distribution is shown in Fig. 3b, which exhibits
a small but significant (at the $3\sigma$ level) peak within 5 arcmin
of the centre of the CCD
survey and $\gamma^{2}$ Vel. The point closest to $\gamma^{2}$ Vel is
missing because the 30 arcsec
region immediately surrounding $\gamma^{2}$ Vel is swamped by its light
and no accurate photometry was obtained there.
In a similar fashion
we can show that the X-ray sources are also marginally concentrated
toward the centre of the field even after correction for the PSPC
vignetting function. 

There is thus some evidence that the PMS stars and
$\gamma^{2}$ Vel are spatially correlated, although this does not rule
out a chance alignment of $\gamma^{2}$ Vel with a background cluster of
low mass stars in Vela OB2. In particular, it is possible that
any concentration we see could be associated with $\gamma^{1}$ Vel, a
B2III, single lined spectroscopic binary which is a
common proper motion companion to $\gamma^{2}$ Vel. $\gamma^{1}$ Vel
is also a likely member of the Vela OB2 association, has a
spectroscopic parallax of $\simeq450$\,pc and is located only 41
arcsecs at a PA of 220$^{\circ}$ from $\gamma^{2}$ Vel.

\subsection{Are the PMS stars and $\gamma^{2}$ Vel at the same
distance?}

The isochrones plotted in Fig.2 come from the D'Antona \& Mazzitelli
(1997) low mass evolutionary models. We have converted from bolometric
luminosity and effective temperature to magnitude and colour using
empirical bolometric corrections as a function of colour and a
colour-effective temperature relationship derived by forcing the well
studied low mass stars in the Pleiades cluster to fit a
125\,Myr isochrone (see Jeffries \& Tolley 1998). 
The assumed distance and reddening are those appropriate for
the Vela OB2 association of 410\,pc and $E(V-I)\simeq 0.06$. It is
clear that if the PMS stars are at the mean distance of the Vela OB2
association they appear to be about $4\pm2$\,Myr old (taking into
account the likely presence of unresolved binary systems).  If however,
the PMS association were at the Hipparcos distance to $\gamma^{2}$ Vel,
the isochrones would be shifted upwards by 1 magnitude. In that case we
would deduce that the PMS population had an age $>10$\,Myr.  The age of
$\gamma^{2}$ Vel, based upon the mass of the O star ($\sim
30$M$_{\odot}$) at the Hipparcos distance, is less
than 5\,Myr (Schaerer et al. 1997; de Marco \& Schmutz 1999). Thus if
the Hipparcos distance is adopted, the PMS stars and $\gamma^{2}$ Vel
cannot be at the same distance {\em and} coeval. 
However, {\em if} $\gamma^{2}$ Vel were at the mean distance of
the Vela OB2 association, it would be more massive (see below), 
slightly younger and could easily be coeval with the PMS population.
An age range of 2-6\,Myr would be compatible with distances between 490
and 360\,pc.

\subsection{How far away is $\gamma^{2}$ Vel?}

The evidence that $\gamma^{2}$ Vel is as close as 258\,pc from the
Hipparcos parallax should be treated with some caution. de Zeeuw et
al. (1999) examine the high mass membership of the Vela OB2 association
on the basis of both proper motions and parallaxes. The association is
well defined by proper motions and has an angular radius of about 6
degrees. The {\em mean} parallax from 93 members corresponds to
$410\pm12$\,pc. The parallax dispersion can be quite well modelled as a
gaussian with a $\sigma=0.68$ mas, such that $\sim95\%$ of the members appear to
lie between 260\,pc and 900\,pc. The dispersion is reasonably
consistent with the errors on the individual points and not
inconsistent with the idea that all the stars are at nearly {\em the
same} distance. Indeed, if we were to assume that the front to back
size of Vela OB2 were similar to its diameter on the sky, then all the
stars should be contained within $\pm40$\,pc.  We would therefore
interpret the parallax to $\gamma^{2}$ Vel simply as a $\sim2\sigma$
deviation and that its distance was $410\pm40$\,pc.

The companionship of $\gamma^{1}$ Vel is also evidence for a distance
closer to the mean Vela OB2 distance. Although these objects have a
common proper motion, they were too close together on the sky for Hipparcos
to obtain independent parallaxes. The distance to
$\gamma^{1}$ Vel from its spectral type and photometry is almost
certainly 400-500\,pc (e.g. Abt et al. 1976; Hern\'{a}ndez \& Sahade 1980).
The idea that $\gamma^{2}$ Vel is a foreground object randomly placed
within 1 arcminute of another bright member of the Vela OB2 association
has a probability of only $\sim10^{-3}$. There is also reasonable
evidence that the radial velocities of the two systems are
similar. Hern\'{a}ndez \& Sahade (1980) quote $9.7\pm 1.0$\kms\ for
$\gamma^{1}$ Vel and Niemel\"{a} \& Sahade (1980) give $12\pm1$\kms\ for
$\gamma^{2}$ Vel, although it is likely that the accuracy of the latter
result is exaggerated (Schmutz et al. 1997). 

\subsection{Can the PMS stars be at a range of distances?}

Preibisch \& Zinnecker (1999) observed a very similar PMS
CMD in the Upper Sco OB association, with a similar vertical scatter
of about $\pm 0.6$ mag about the isochrones. They showed that if one
takes into account unresolved binaries, photometric errors and allowed
a $\sim10\%$ range in distance, that the scatter around the PMS
isochrones was exactly as expected for a coeval population at $\sim5$\,Myr. 
We therefore similarly conclude
that the CMD in Fig. 2 shows no evidence for a 
large spread in {\em either} the age or distance
of the PMS population we have found. In particular, we can rule out any
distance modulus spread in a coeval population that is larger than a few tenths
of a magnitude, or any age spread in a co-spatial population of more than a Myr
or so (for a mean age of $\sim4$\,Myr). Thus unless there is a
conspiracy to place older stars closer to us, the PMS association seems
likely to have a relatively narrow spread around an age and distance that are
incompatible with the deduced age for $\gamma^{2}$ Vel and its
Hipparcos distance.

\subsection{Is a larger distance to $\gamma^{2}$ Vel consistent with
its physical properties?}

Our findings challenge the conclusions of several recent papers which
use the Hipparcos parallax of $\gamma^{2}$ Vel and its error to derive:
the absolute magnitude of the system and its components; system masses
from the interferometric binary separation of Hanbury-Brown (1970); the
O star luminosity, mass and age from stellar evolution models and hence
the orbital inclination and further mass estimates from radial velocity
curves (see van der Hucht et al.  1997; Schaerer et al. 1997, Schmutz
et al. 1997, de Marco \& Schmutz 1999).  A distance as large as 410\,pc
for $\gamma^{2}$ Vel significantly changes the system parameters
deduced in these papers.  The system luminosity increases by a factor
2.5. The effective temperature and luminosity deduced for the O star
would then give it a mass $>40$\msun\ and an age $<3$\,Myr, compared
with the values of 30\msun\ and 3.6\,Myr quoted by de Marco \& Schmutz
(1999). At this larger distance, an age of 2-3\,Myr could be compatible
with the low mass PMS stars we have found. The absolute magnitude of
the O star would decrease to $-6.0\pm0.3$ (van der Hucht et al. 1997),
which argues for a supergiant rather than a giant classification. Van
der Hucht et al. comment that this would be in better agreement with
published spectra of Conti \& Smith (1972) and Niemel\"{a} \& Sahade
(1980).  As the mass ratio from the radial velocity curves is fixed,
the WR mass increases by a similar fraction. The total system mass,
based on a binary separation of $4.3\pm0.5$ mas, increases from
$30\pm10$\msun\ to $120\pm40$\msun\ (now comfortably exceeding the
minimum mass from radial velocity curves -- Niemel\"{a} \& Sahade 1980)
and the binary inclination is reduced to around $50^{\circ}$ to explain
the radial velocity curves.

\section{Conclusions}

From our discussion there seem to be two possible scenarios. (1) That
the PMS stars are approximately at the same distance and age as
$\gamma^{2}$ Vel, and that this distance places $\gamma^{2}$ Vel within
the Vela OB2 association at 360-490\,pc.  (2) That the PMS stars are
part of the Vela OB2 association, possibly surrounding $\gamma^{1}$
Vel, but that $\gamma^{2}$ Vel is an isolated foreground object with no
surrounding low mass stars at a similar age. We believe that (1) is
{\em far} more plausible than (2) because of the dispersion in the Vela
OB2 Hipparcos parallaxes and the likely association of $\gamma^{1}$ and
$\gamma^{2}$ Vel. Recently, the idea that $\gamma^{2}$ Vel could form
in isolation without accompanying low mass stars has also been
challenged by the near IR detection of a K-type PMS companion only 4.7
arcsec distant (Tokovinin et al. 1999).

If the low mass PMS stars we have found are truly in the vicinity of
$\gamma^{2}$ Vel, they represent an exciting opportunity to explore the
influence of adjacent high mass loss stars and ionizing UV radiation
fields on the mass function and circumstellar disc lifetimes of low
mass stars. It will be interesting to compare the frequencies of
T-Tauri discs around these stars with the frequencies found in
T associations and OB associations with similar ages.
The PMS stars in Fig.2 have masses, found from the D'Antona
\& Mazzitelli (1997) models, down to (an age dependent) mass of 
about 0.15\msun. The mass function will be addressed when we
have a better census of the association membership.

\section*{ACKNOWLEDGMENTS}

This research has made use of {\em ROSAT} data obtained from the
Leicester Database Archive Service at the Department of Physics and
Astronomy, Leicester University, UK.  The Cerro Tololo Interamerican
Observatory is operated by the Association of Universities for Research
in Astronomy, Inc., under contract to the US National Science
Foundation. TN was supported by a UK Particle and Physics and
Astronomy Research Council (PPARC) Advanced Fellowship.  SH was
supported by a Nuffield Foundation Undergraduate Research Bursary
(NUF-URB98). EJT is a PPARC postdoctoral research associate.  MK was
supported by an undergraduate research bursary from Keele University.
Computational work for this paper was performed on the Keele node of
the Starlink network funded by PPARC.

\label{lastpage}

\end{document}